\documentclass{article}
\usepackage[ a4paper,
 total={170mm,257mm},
 left=20mm,
 top=20mm]{geometry}
\usepackage{afterpage}

\usepackage{graphicx}

\usepackage{pgfplots}
\usepackage{pgfplotstable}
\usepackage{longtable}
\usepackage{multicol}      
\twocolumn
\usepackage{flushend} 
\newcommand{\fscale}{1} 
\usepackage{caption}

\usepackage{hyperref}
\usepackage{siunitx}
\usepackage{amssymb} 
\usepackage{amsmath}

\usepackage{multirow}
\usepackage{kantlipsum}
\usepackage{dblfloatfix} 

\pgfkeys{/pgf/number format/.cd,sci e}
\newsavebox\ltmcbox
\usepackage{authblk} 

\begin{document}

\title{Combining neutrino experimental light-curves for pointing to the next Galactic Core-Collapse Supernova}
\author{A. Coleiro$^{1}$, M. Colomer Molla*$^{1,2}$, D. Dornic$^{3}$, M. Lincetto$^{3}$, V. Kulikovskiy*$^{4}$}

\affil{\small{$^{1}$Universit\'e de Paris, CNRS, Astroparticule et Cosmologie, F-75013 Paris, France\\
$^{2}$IFIC - Instituto de F\'isica Corpuscular (CSIC - Universitat de Val\`encia) c/ Catedr\'atico Jos\'e Beltr\'an, 2 E-46980 Paterna, Valencia, Spain\\
$^{3}$Aix Marseille Univ, CNRS/IN2P3, CPPM, Marseille, France\\
$^{4}$INFN - Sezione di Genova, Via Dodecaneso 33, 16146 Genova, Italy}}



\twocolumn[
\begin{@twocolumnfalse}
\maketitle
\begin{abstract}
The multi-messenger observation of the next galactic core-collapse supernova will shed light on the different physical processes involved in these energetic explosions. Good timing and pointing capabilities of neutrino detectors would help in the search for an electromagnetic or gravitational-wave counterparts.

An approach for the determination of the arrival time delay of the neutrino signal at different experiments using a direct detected neutrino light-curve matching is discussed. A simplified supernova model and detector simulation are used for its application. The arrival time delay and its uncertainty between two neutrino detectors are estimated with chi-square and cross-correlation methods. The direct comparison of the detected light-curves offers the advantage to be model-independent.

Millisecond time resolution on the arrival time delay at two different detectors is needed. Using the computed time delay between different combinations of currently operational and future detectors, a triangulation method is used to infer the supernova localisation in the sky. The combination of IceCube, Hyper-Kamiokande, JUNO and KM3NeT/ARCA provides a 90\% confidence area of 140$\pm$20\,deg$^2$. These low-latency analysis methods can be implemented in the SNEWS alert system.
\end{abstract}
\vspace{1cm}
\end{@twocolumnfalse}
]

\section{Introduction}

The two dozen of neutrinos observed from the SN1987A explosion at the Large Magellanic Cloud indicate that Core-Collapse Supernovae (CCSN) are sources of 1--100~MeV neutrinos, with a total energy emitted on the order of $3\times10^{53}$\,erg. In such explosive phenomena, 99\% of the total gravitational binding energy of the stellar collapse is released through neutrinos~\cite{Janka}. Neutrinos leave the stellar core several hours before photons. The recent multi-wavelength observations of cataclysmic transient events in the Universe, and in particular the multi-messenger detection of the binary neutron star merger GW170817~\cite{GW170817}, have shown the crucial importance of combining neutrinos, gravitational waves and electromagnetic signals to unveil the mechanisms driving these astrophysical events. Therefore, fast and precise direction reconstruction of the neutrino flux, as well as of the arrival signal time, are important for an efficient multi-messenger follow-up. 

In this work, several water Cherenkov detectors are considered: the underground detectors, Super-Kami\-okande~\cite{SKgeneral} and Hyper-Kamiokande~\cite{HKgeneral}, and the high-energy neutrino telescopes, IceCube~\cite{ICgeneral} and KM3\-NeT \cite{LoIKM3NeT}. The main interaction channel in water Cherenkov detectors is inverse beta decay (IBD) of electron anti-neutrinos ($\overline{\nu}_{e}$) on proton targets. The positrons angular distribution is slightly forward-peaked and high energy events can be selected in order to exploit this directionality~\cite{Vissani}. Because of the weak anisotropy, pointing to CCSN with IBD is difficult in water Cherenkov detectors and it is more promising for liquid scintillator detectors~\cite{LiquidScin}. The JUNO scintillator detector is sensitive to both IBD and elastic scattering (ES)~\cite{JUNO_SN_sensitivity}. Moreover, for the IBD channel JUNO offers the ability to identify positrons and neutrons and reconstruct their positions. The direction along the line connecting the positions of both products can be used to infer the neutrino direction~\cite{Vogel}. Detectors sensitive to ES interactions can provide information on the CCSN localisation. Nowadays, Super-Kamiokande is the only running detector with enough sensitivity to the ES channel to be able to point by itself to the source~\cite{SK2, SKpointing}.

By exploiting the time delay between the arrival of the signal at different detector sites, it is possible to use a triangulation method to infer the source localisation on the sky~\cite{Burrows,Beacom,2013t,Vedran,Kate,Hansen}. In the triangulation method proposed in~\cite{Burrows}, the uncertainty on the arrival time delay between a pair of detectors is inferred from the number of the detected neutrinos in the bulk of the emission and its duration. In~\cite{Beacom} arrival time measurements at each detector from the bulk of the events and from the first events are described and the maximum precision is estimated using Rao-Cramer theorem with the conclusion that low event statistics detectors available at that time were not useful for triangulation. A direct light-curve comparison between detectors with a Kolmogorov test was also mentioned among the methods. In~\cite{2013t}, the triangulation method is revisited and a rough estimate of the arrival time uncertainty for each detector is computed assuming a generic neutrino light-curve with an exponential rise. In~\cite{Vedran}, a more detailed light-curve template is used. The use of the time delay estimate between the first detected events in each experiment is proposed in~\cite{Kate}. The latter method can be implemented for real-time CCSN localisation. In order to reach a good accuracy, an almost background-free experiment is required. This implies that the use of large volume Cherenkov neutrino telescopes, as IceCube and KM3NeT, is more difficult. Timing with the first events is earlier proposed in~\cite{Hansen} together with the exponential rise fit.

In this work, a model-independent approach that relies on matching the detected neutrino light-curves is elaborated. Such an approach requires data sharing between the detectors. Communication like this could be enabled via the SNEWS global network~\cite{SNEWS}. The elaborated methods are tested on a parametric neutrino flux function and with a simplified detector description to tune the method performance and test the absence of intrinsic biases. Triangulation is demonstrated for a benchmark CCSN at 10~kpc distance. The provided results are intended for a rough estimation of the method performance and as benchmark results in case the methods are reproduced by the readers. The codes for the detected neutrino light-curves simulation, their matching and skymaps creations are publicly available~\cite{codes}.

The paper is organized as follows. The simulation framework to model CCSN neutrino detection rate at different detectors is introduced in Section~\ref{sec:sim}. In Section~\ref{s:method}, the statistical methods used to estimate the time difference of the neutrino arrival at the different sites are described. In Section~\ref{s:results}, the arrival time delay resolution for different detector pairs is discussed. Different combinations of three and four detectors are then used to evaluate the CCSN localisation uncertainty. Conclusions are provided in Section~\ref{s:concl}.
\section{Simulation}
\label{sec:sim}

The CCSN neutrino detection rates at different observatories are estimated with a simple simulation described in this section. It includes a simplified neutrino luminosity curve and the detector parameters describing their detection efficiency and background.

The main simplifications are the following: time and energy independent detector efficiency, and a steady energy neutrino emission. These approximations are physically motivated for the context of this work, and they allow fast and easily reproducible detected neutrino light-curve simulations, which are required for  testing and optimising the method.

It is known that the detectors efficiency is generally improved with the neutrino energy. Using more realistic detector descriptions should be done together with using detailed CCSN emission models to fully estimate non proportional signal rates in the detectors. Detector limitations such as event pileup and other efficiency variations can further increase the differences between the detector responses. Neutrino flux simulation can be affected by neutrino mixing at the CCSN. Even larger differences between the various detailed models are present for some available progenitors~\cite{Tamborra}. Moreover, the distance to the next Galactic CCSN is unknown, and this will affect the absolute flux scale.

\subsection{Supernova neutrino fluxes}
\label{fluxes}

The all-flavour neutrino total luminosity is set to $L_{\nu}^{0}=3\times10^{53}$\,erg, as in~\cite{JUNO_SN_sensitivity}. It is equally divided among all neutrino flavours and between neutrinos and anti-neutrinos. This is consistent with the more accurate estimate of the $\bar\nu_e$ fraction of about 
0.14 that includes oscillations in the supernova mantle~\cite{Vissani}.

\begin{figure}[h] 
\centering
\includegraphics[width=\fscale\linewidth]{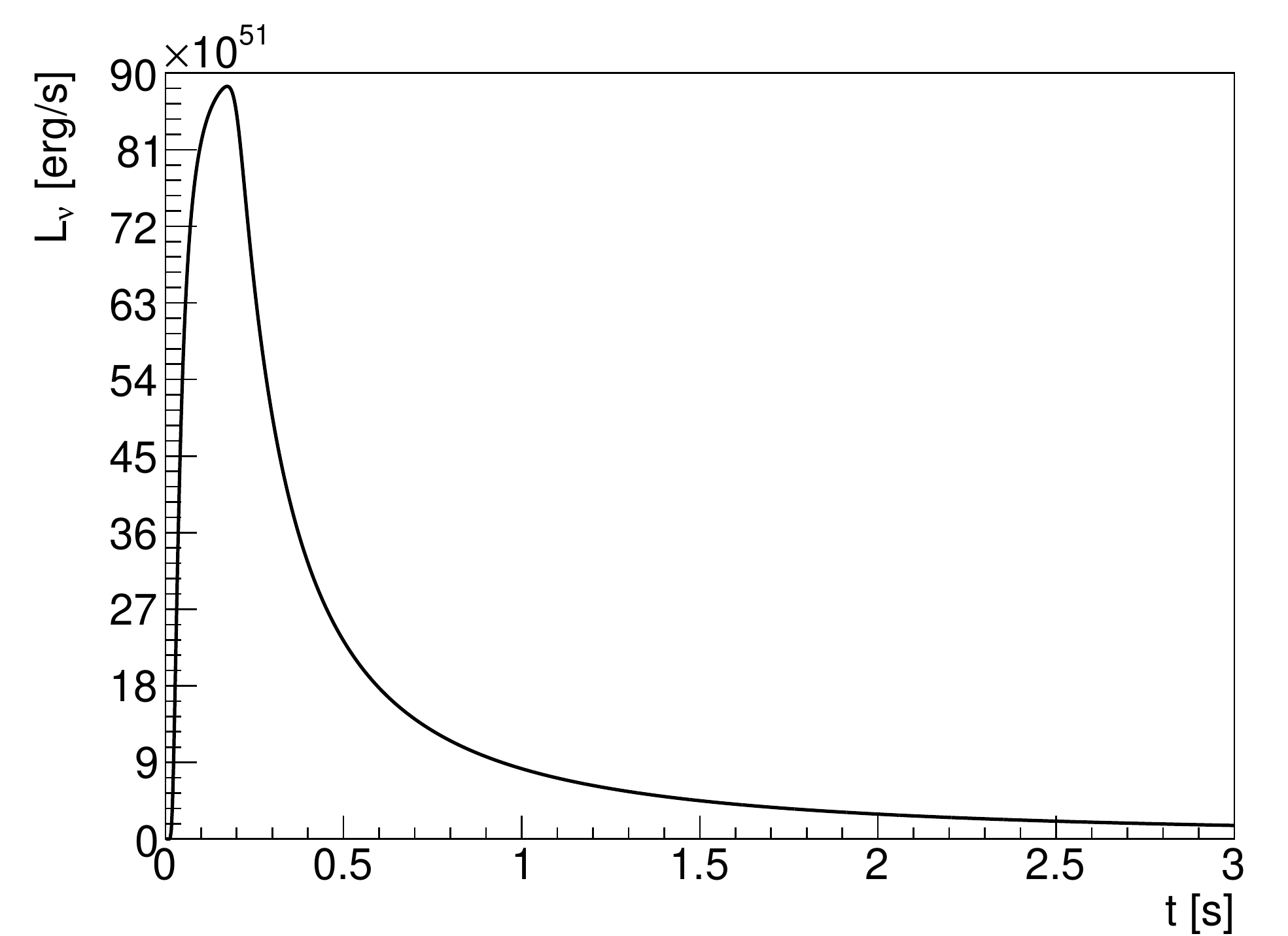} 
\caption{Time evolution of the neutrino luminosity in the considered simplified model.}
\label{fig:fluxes}
\end{figure}

The time evolution of the neutrino luminosity is taken from~\cite{simplifiedflux} and it is described for $\bar\nu_e$ as:
\begin{equation}
L_{\nu}(t) = \frac{L_{\nu}^{0}}{6} \frac{1}{\mathcal{N}}\frac{e^{-(t_a/t)^{n_a}}}{[1+(t/t_c)^{n_p}]^{n_c/n_p}} \: ,
\end{equation}
where $\mathcal{N}$ is used to normalise the function as follows:
\begin{equation}
\mathcal{N}=\int_0^\infty\frac{e^{-(t_a/t)^{n_a}}}{[1+(t/t_c)^{n_p}]^{n_c/n_p}}\,dt \: .
\label{eq:norm}
\end{equation}
For these studies the chosen parameters are: $t_a=0.035$\,ms, $t_c=0.2$\,ms, $n_a=2$, $n_p=20$ and $n_c=1.5$. The light-curve simulated using these parameters approximately follows the predicted accretion phase $\bar\nu_e$ luminosity in~\cite{Tamborra}. The chosen value of $n_c=1.5$ guarantees the convergence of the luminosity integral. The luminosity curve with the assumed parameters is shown in Figure~\ref{fig:fluxes}.

The energy distribution can be described by a quasi-thermal distribution \cite{Tamborra}:
\begin{multline}
f_E(E_\nu) = \\ \frac{1}{\Gamma(1+\alpha)} \left( \frac{1+\alpha}{\tilde{E_{\nu}}}\right)^{1+\alpha}E_{\nu}^\alpha  \exp\left(-\frac{(\alpha+1)E_{\nu}}{\tilde{E_{\nu}}}\right) \: ,
\end{multline}
which depends on the average neutrino energy, $\tilde{E_{\nu}}$, and the spectral pinching shape parameter, $\alpha$. Both parameters are generally varying with time. The typical range of the pinching shape parameter is $2\lesssim\alpha\lesssim5$~\cite{Tamborra}.

Only the variation in time of the detection rates is relevant for this study. Therefore, the simulation follows the luminosity time evolution with a steady energy neutrino spectrum with $\tilde{E_{\nu}}=14$\,MeV and $\alpha=3$. The neutrino luminosity can be converted into a flux dividing it by the average neutrino energy. Assuming an isotropic neutrino emission from a source at a distance $d$, the differential neutrino flux is:
\begin{equation}
\frac{d\Phi}{d E_\nu}(E_\nu,t)= \frac{1}{4 \pi d^{2}} \frac{L_{\nu}(t)}{\tilde{E_{\nu}}} f_E(E_\nu) \: .
\end{equation}

\subsection{CCSN neutrino detection rates}

\begin{table*}[!b]
\caption{Simplified detector characteristics used as input for the simulation: effective mass in water equivalent units and total background rate.}
\label{tab:detectors}
\begin{center}
\begin{tabular}{|c|c|c|c|c|c|c|}\hline   
 & {\bf IceCube} & {\bf ORCA} & {\bf ARCA} & {\bf Super-Kamiokande} & {\bf Hyper-Kamiokande} & {\bf JUNO} \\ \hline
$M_\mathrm{eff}$ [kton] & 3500 & 90 & 180 & 22.5 & 560 & 22.5 \\ \hline
$R_\mathrm{bg}$ [Hz] & $\sim$3e6 & $\sim$1e6 & $\sim$2e6 & 0 & 0 & 0 \\ \hline
Reference & \cite{ICSN} & \cite{ICRCprocSNKM3,MUONDEPTH} & \cite{ICRCprocSNKM3,MUONDEPTH} & \cite{Migenda, SK, SK2} & \cite{Migenda, HK} & \cite{Juno} \\ \hline
\end{tabular}
\end{center}
\end{table*}

As it is the main channel for the considered water detectors, only the inverse beta decay interaction channel is simulated.

The instantaneous event rate in the detector is estimated as the product of the differential neutrino flux, $\frac{d\Phi}{d E_\nu}$, the IBD cross-section, $\sigma_\mathrm{IBD}$~\cite{Vissani}, the number of targets per unit volume, $2 \frac{\rho N_A}{\mu_{H_2O}}$, the detection efficiency, $\epsilon_{det}$, and the detector volume $V_\mathrm{det}$:
\begin{multline}
R_\mathrm{sig}(t) =
\int_0^\infty \frac{d\Phi}{d E_\nu}(E_\nu,t) \sigma_\mathrm{IBD}(E_\nu) \\ \times2\frac{\rho N_\mathrm{A}}{\mu_{H_2O}} \epsilon_\mathrm{det}(E_\nu)  V_\mathrm{det} \, d E_\nu \: .
\label{eq:Rsig}
\end{multline}
The factor 2 comes from the two hydrogen atoms in a water molecule, $\mu=18$\,g/mol is the molar mass of water, $\rho$ is the density of water and $N_\mathrm{A}$ is the Avogadro number.

The detector properties are converted into the detector effective mass, $M_\mathrm{eff}$, in the following way:
\begin{equation}
M_\mathrm{eff}(E_\nu) = \rho \epsilon_\mathrm{det} (E_\nu) V_\mathrm{det} \: .
\end{equation}

The detection efficiency depends on the neutrino energy, or more precisely, on the energy of the detectable interaction products. A constant efficiency is assumed above an energy threshold, $E_\nu^\mathrm{min}$. Since this value is around the Cherenkov threshold, and below the energy range where supernova neutrinos are expected, $E_\nu^\mathrm{min}=0$ is set for simplicity. This assumption removes the energy dependence of the $M_\mathrm{eff}$. The detection rate is calculated as:
\begin{equation}
\begin{split}
R_\mathrm{sig}(t) &= M_\mathrm{eff} \frac{L_\nu(t) }{\tilde{E_\nu}} \frac{1}{4 \pi d^{2}} \\ 
&\times\int_{E_\nu^\mathrm{min}}^\infty f_E(E_\nu) \sigma_\mathrm{IBD}(E_\nu) 2\frac{N_\mathrm{A}}{\mu_\mathrm{H_2O}} \, d E_\nu \\
&= M_\mathrm{eff} L_\nu(t) I \:,
\label{eq:rates}
\end{split}
\end{equation}
where the simplified conversion parameter, $I$, is the same for all the detectors. For a CCSN at 10\,kpc, this factor becomes: 
$I\approx4.3\,\mathrm{erg}^{-1}\mathrm{kton}^{-1}$.

The rates for future scintillator or other non-water detectors can be calculated using Equation~\ref{eq:rates}, where $M_\mathrm{eff}$ is estimated in water equivalent units and $N_\mathrm{target}$ is the total number of targets in the detector:
\begin{equation}
M_\mathrm{eff}^\mathrm{w.e.}=\epsilon_\mathrm{det} N_\mathrm{target} \left( \frac{2N_\mathrm{A}}{\mu_\mathrm{H_2O}} \right)^{-1} \: .
\end{equation}

\subsection{Simplified detector model} 
The detector model is described by two parameters: the supernova detection effective mass, $M_\mathrm{eff}$, and the background rate, $R_\mathrm{bg}$. The signal rates are estimated for each detector from Equation~\ref{eq:rates}. Both the signal and the background rates are translated into an expected number of events per time bin. In order to simulate experimental fluctuations in the detected neutrino light-curve, the number of events in each time bin is sampled assuming a Poisson distribution.

For the Super-Kamiokande and Hyper-Kamiokande detectors, the effective masses are taken from~\cite{Migenda, SK, HK}. For JUNO, the expected number of proton targets is $N_\mathrm{target}=1.5\times10^{33}$~\cite{Juno}, similar to Super-Kamiokande. The same effective mass in water equivalent units, $M_\mathrm{eff}=22.5$\,kton, is used for both detectors. In this study, the background rates are negligible for Super-Kamiokande, Hyper-Kamiokande~\cite{SK2} and JUNO~\cite{Juno}.

The assumed IceCube detector effective volume is 3.5\,Mton~\cite{ICSN}. The detector consists of 5160 optical modules, each containing one 10-inch photomultiplier. The background rate per optical module is about 300~Hz, taking an average value between standard and high efficiency optical modules.

The optical module of KM3NeT consists of 31 3-inch photomultipliers. Using nanosecond scale coincidences between the photomultipliers on the same optical module, the variable background from the bioluminescence can be suppressed~\cite{ICRCprocSNKM3}. The remaining background contribution is mostly coming from $^{40}$K decays in sea water and it amounts to a total rate of $R_{OM}^{K40}\sim$500~Hz per optical module~\cite{MUONDEPTH}.  The effective mass of the KM3NeT detectors is estimated as $M_\mathrm{eff}\approx180$\,kton for KM3NeT/ARCA and 90~kton for KM3NeT/ORCA from~\cite[Fig. 4, left]{ICRCprocSNKM3}. The ARCA and ORCA detectors consist of 4140 and 2070 optical modules, respectively. The coincidence selection translates into a lower effective CCSN detection volume for KM3NeT/ARCA compared to IceCube, even if both have an instrumented volume of $\sim$1~km$^{3}$.

\begin{figure*}[!h]
\centering
\includegraphics[scale=0.42]{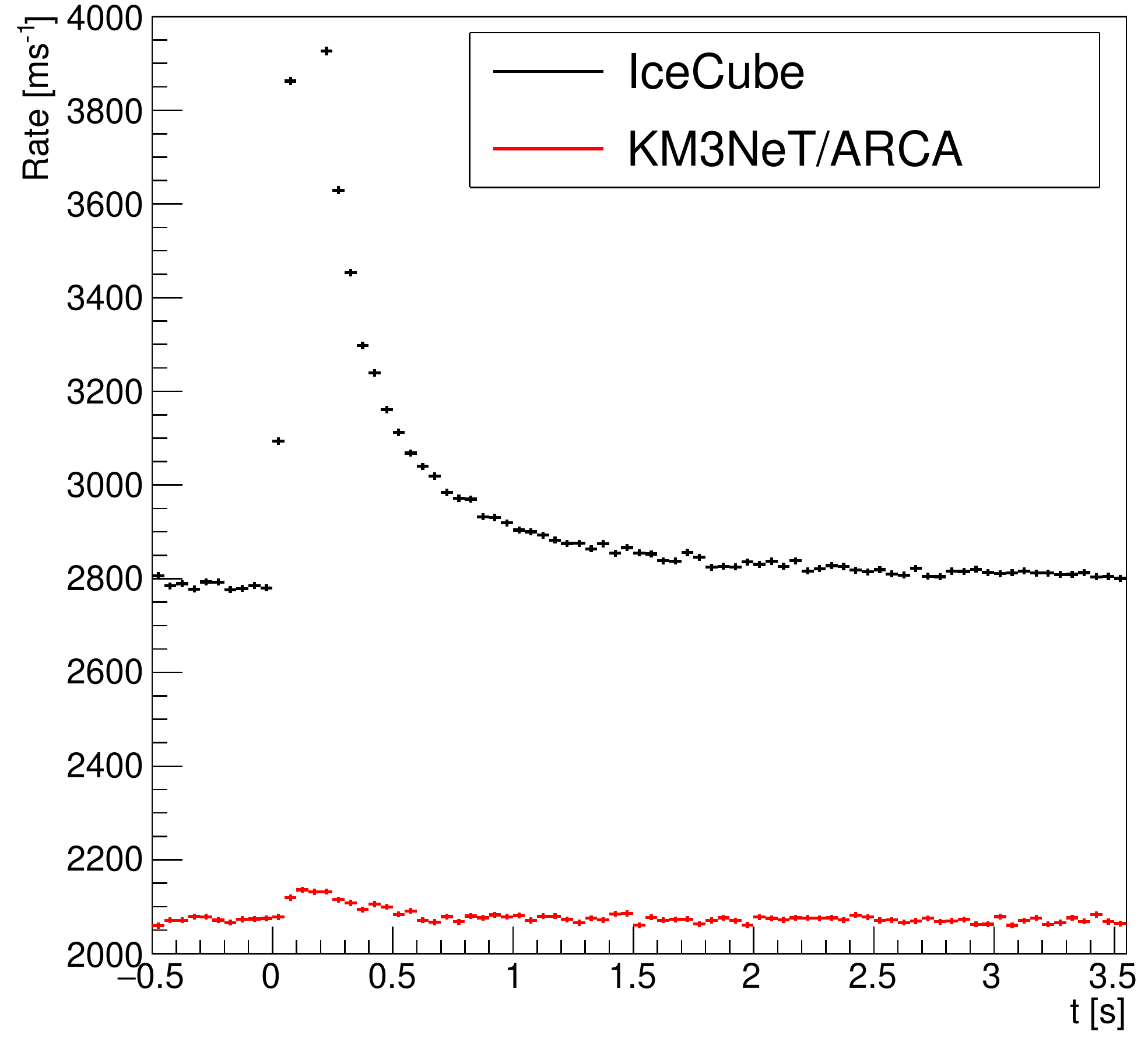}
\includegraphics[scale=0.42]{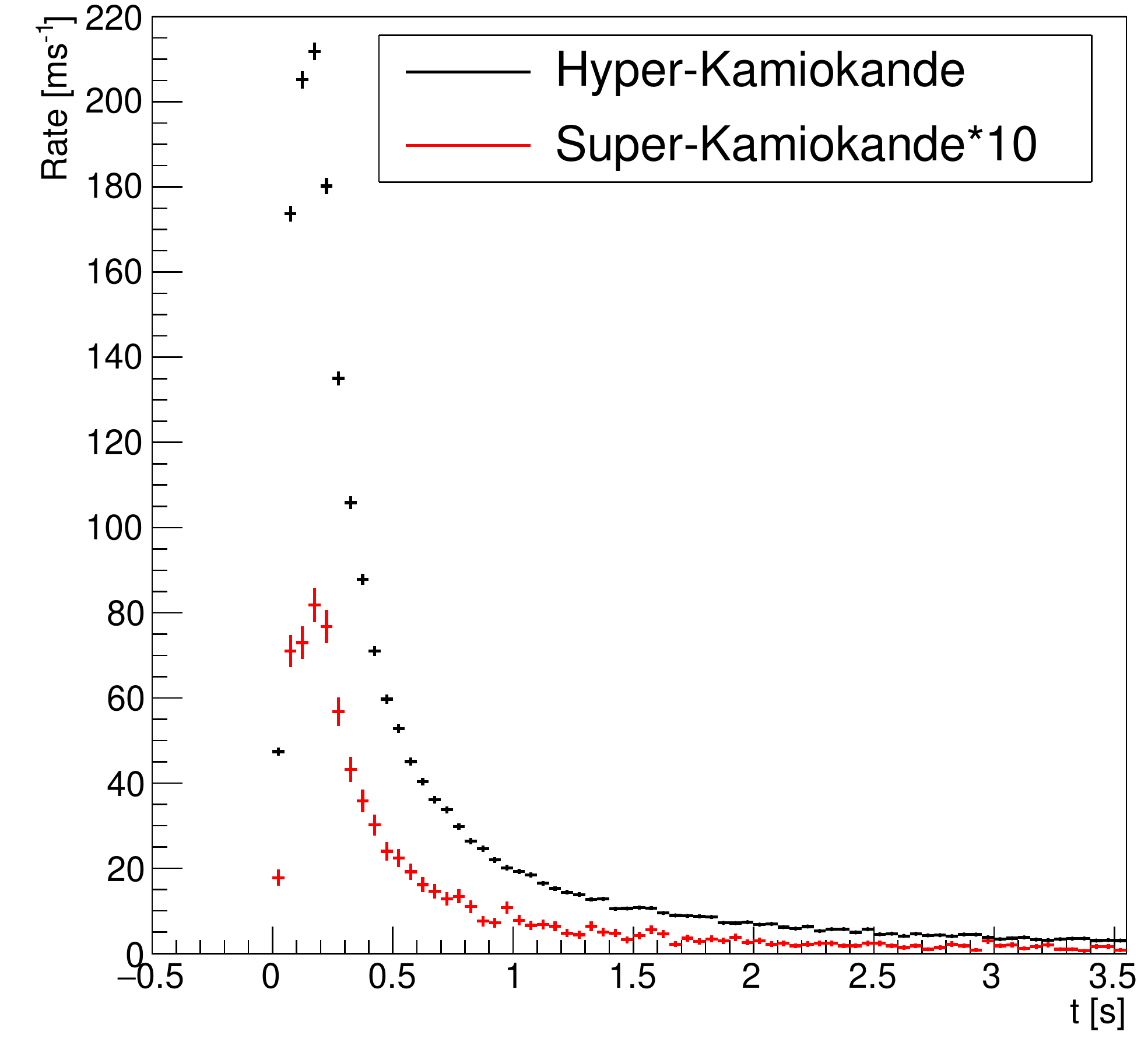}
\caption{Simulated neutrino light-curves with 50\,ms binning, showing the event rates for IceCube (black) and ARCA (red) on the left and Hyper-Kamiokande (black) and Super-Kamiokande, zoomed vertically 10 times for visual purposes (red) on the right. Error bars for each bin correspond to the original number of events in  each bin scaled accordingly.}
\label{fig:LC}
\end{figure*}

The effective mass and the background rate for the different detectors used in this work are summarised in Table~\ref{tab:detectors}. Some examples of the simulated detector light-curves are provided in Figure~\ref{fig:LC}.

\section{Description of the methods}
\label{s:method}
Two methods to estimate the time difference of the neutrino arrival times between two detectors have been investigated: the chi-square (Section~\ref{s:chi2}) and the cross-correlation (Section~\ref{s:cc}). Performance estimation with a large number of simulations is described in Section~\ref{s:methperfomance} using a simplified CCSN model. A bootstrapping procedure using detected light-curve is introduced in Section~\ref{s:bootstrapping}.
The triangulation technique for the source localisation is reviewed in Section~\ref{s:triang} and the methods for the triangulation performance estimation are described in Section~\ref{s:triangperformance}.

\subsection{Chi-square method}
\label{s:chi2}

A method based on $\chi^{2}$ minimisation has been developed and tested to infer the delay between two light-curves. This technique is one of the traditional ways to verify the compatibility of two distributions and for parameter estimation. In this case, the estimated parameter is the time delay between the two detected neutrino light-curves.



Assuming a normal distribution of the number of events in each time bin the $\chi^{2}$ expression is defined as follows:
\begin{multline}
\chi^{2} (\tau) = \\ \sum_{t_i=t_\mathrm{min}}^{t_\mathrm{max}}  \frac{( (n_{t_i-\tau} - m_{t_i}) - E(n_{t_i-\tau} - m_{t_i}) )^{2} }{V(n_{t_i-\tau} - m_{t_i})} \: ,
\end{multline}
where $\tau$ is the time shift between the detected neutrino light-curves, $n_{t_i-\tau}$ is the number of observed events by the first detector in the time bin $t_i-\tau$, $m_{t_i}$ is the number of observed events by the second detector in the time bin $t_i$, $E(n_{t_i-\tau} - m_{t_i})$ and $V(n_{t_i-\tau} - m_{t_i})$ are the expectation value and the variance of the difference in the number of events, respectively. For two normal distributions, the variance of the difference between the number of events corresponds to the sum of their squared standard deviations:
\begin{equation}
\begin{split}
V(n_{t_i-\tau}-m_{t_i})&=V(n_{t_i-\tau})+V(m_{t_i}) \\
&=(\sigma_{n_{t_i-\tau}})^2+(\sigma_{ m_{t_i}})^2 \: .
\end{split}
\end{equation}

The value of $\tau=T_0^\mathrm{match}$ obtained at the $\chi^{2}$ minimum provides the best estimate of the true time delay between the two detectors, $T_0$.

A different expression of the $\chi^{2}$ was used in \cite{Vedran}. It assumes that the number of events in each time bin follows a Poisson distribution with a mean value equal to the average expected number of events computed from a CCSN neutrino emission model. This Poisson $\chi^{2}$ expression cannot be used to compare two experimental light-curves since it requires a known expectation value. Since the number of detected events follows a Poisson distribution, the bin size should be optimized in our method to make the Gaussian approximation valid.

If the light-curves are normalised in a way that signal and background expectations are almost identical for both detectors, then the expectation value $E(n_{t_i-\tau} - m_{t_i})$ can be considered null for any $t_i$ when $\tau$ corresponds to the true shift. To achieve such normalisation, the mean detector background is subtracted from the light-curve so that all combined detectors have a common baseline centred at 0. The background expectation value can be taken from a time window in the detected neutrino light-curve where CCSN signal is not present (off-signal zone) or it can be provided by each experiment based on a longer detector monitoring. In this study, an off-signal zone of 1\,s duration is chosen for the background estimate. In order to account for the different detector efficiencies, a normalisation to the detector effective mass is applied. Alternatively, this effect can be taken into account by normalising each light-curve after background subtraction to have unit integral. Note that such light-curve normalisation assumes a constant expectation value of the background rates and detector efficiencies. This is in perfect agreement with the simplified simulations, as described earlier. In the real case, this is generally not true. For example, a neutrino energy spectrum varying in time and an energy dependent detector efficiency may affect the results. The described normalisation can be improved to account for these effects by using time-dependent efficiency and background rate expectations. Introducing the time dependent efficiency correction may involve the convolution of the detector efficiency as a function of the energy with the neutrino emission spectrum varying in time. The detailed neutrino emission spectrum differs amongst the models so this may bring a bias to a such sophisticated correction.

The signal arrival time is not known \textit{a priori} in this model independent analysis. Therefore, a reference time is evaluated to define the window for the $\chi^{2}$ calculation. The time window of [--300, 300]\,ms centred at the maximum of the detected light-curve from one of the detectors is chosen for this analysis. The interval covers the transition between the background and the CCSN neutrino signal as well as the accretion phase, for which most of the neutrino emission is expected~\cite{Tamborra}. Using a time window too long may lead to a degradation of the performance since the optimisation will be more sensitive to background fluctuations. The $\chi^{2}$ calculation window is fixed with respect to the less performing detector to preserve its background statistics during $\tau$ scan. The choice of the fixed detector and the detector for the reference time definition is done to minimize the estimated time delay uncertainty. The procedure for the uncertainty estimation is given in Section~\ref{s:methperfomance}.

The detected light-curves are provided as histograms with a fixed bin size. A bin width of 0.1\,ms is chosen, and the same value is used for the time shift step $\tau$, in order to reach $\mathcal{O}$(ms) required resolution. The numbers of events, $n_{t_i-\tau}$ and $m_{t_i}$, are calculated by summing the events of contiguous bins. The resulting effective bin size is optimised for each detector pair to reach the minimum of arrival time delay uncertainty, $\delta t$. Steps of $t_{i+1}-t_i$ can be optimised in order to increase the calculation speed.

\subsection{Normalised cross-correlation}
\label{s:cc}
The cross-correlation can be used for matching two sets of time series~\cite{FermiCC} or for matching a time series with a template model~\cite{bookcc}. The discrete cross-correlation is defined as:
\begin{equation}
\label{eq:cc}
\mathcal{C}(\tau)=(n \star m) = \frac{1}{N} \sum_{t_{i}=t_\mathrm{min}}^{t_\mathrm{max}} n_{t_{i}} m_{t_{i}-\tau} \: ,
\end{equation}
where $n_{t_i}$ and $m_{t_i-\tau}$ are the number of observed events by the first and the second detector in time bins $t_i$ and $t_i-\tau$, respectively. $N$ is the number of bins in the search window $[t_\mathrm{min},t_\mathrm{max}]$, and $\tau$ is the time delay between the two light-curves. The function will present a maximum at $\tau=T_0^\mathrm{match}$, allowing to estimate the time delay between the two detectors, $T_0$.

In order to account for the different effective masses and background rates of each detector, the zero-norm\-alised cross-correlation is used. Each curve is normalised by subtracting its mean value and scaled by its standard deviation~\cite{FermiCC}, which can be computed in the search window:
\begin{equation}
\tilde n = \frac{\sum_{t_{i}=t_\mathrm{min}}^{t_\mathrm{max}} n_{t_{i}}}{N}, \tilde m = \frac{\sum_{t_{i}=t_\mathrm{min}}^{t_\mathrm{max}} m_{t_{i}}}{N}\: ,
\end{equation}

\begin{align}
\sigma_n &= \sqrt{\frac{\sum_{t_{i}=t_\mathrm{min}}^{t_\mathrm{max}} (n_{t_{i}}-\tilde n)^2}{N-1}}, \\
\sigma_m &= \sqrt{\frac{\sum_{t_{i}=t_\mathrm{min}}^{t_\mathrm{max}} (m_{t_{i}}-\tilde m)^2}{N-1}}\: .
\end{align}

One of the advantages of this method is that fast Fourier transformations can be used to speed up the calculations and the improvement of the response time can be significant for a large number of bins compared to the chi-square method.

\subsection{Procedure for the light-curve matching performance estimation}
\label{s:methperfomance}
Simulations of the different detected neutrino light-curves are performed following the model described in Section~\ref{sec:sim}. An artificial delay of the neutrino signal observed between two sites, $T_0$, is applied to the first light-curve. The result is not expected to change when exchanging the two detectors. The best estimate $T_0^\mathrm{match}$ is inferred with the proposed methods (see Sections~\ref{s:chi2} and \ref{s:cc}). Finally, $T_0^\mathrm{match}$, is compared to the true $T_0$ value.

The distribution of $T_0^\mathrm{match} - T_0$ is built from a large number of realisations of the simulated light-curves. To confirm the absence of systematic effects, the distribution of $T_0^\mathrm{match} - T_0$ should be compatible with a normal distribution with mean zero and standard deviation independent of $T_0$. The width of the distribution provides an estimate of the uncertainty on the $T_0$ measurement, $\delta t$.

\subsection{Bootstrapping parameters tuning}
\label{s:bootstrapping}
The proposed methods present some parameters that can be slightly tuned for different CCSN models, distances to the source and detectors. In most cases, it was verified that the degradation of the time precision is not significant and some common set of parameters can be identified for future combined analysis.

In order to reach the best performance once the supernova is detected, the following bootstrapping procedure can be set up. The detected neutrino light-curve from the best performing detector is used as the model. With this model, the detected neutrino light-curves for all the detectors can be simulated and the procedure to estimate the performance described in Section~\ref{s:methperfomance} can be repeated in order to tune the matching algorithms parameters.


\subsection{The triangulation method}
\label{s:triang}
The time difference between the CCSN neutrino signal arrival at two detectors located at $\vec{r_{i}}$ and $\vec{r_{j}}$ can be expressed as: 
\begin{equation}
t_{ij} = (\vec{r_{i}} - \vec{r_{j}}) \cdot \vec{n} / c,
\label{eq:tij}
\end{equation}
where $\vec{n}$ is the unit vector that indicates the emission direction. This vector is calculated in the geographic horizontal coordinate system from the CCSN right ascension, $\alpha$, declination, $\delta$, and the Greenwich mean sidereal time expressed as angle, $\gamma$:
\begin{multline}
\vec{n}=\\
(-\cos(\alpha-\gamma)\cos\delta,\,-\sin(\alpha-\gamma)\cos\delta,\,-\sin\delta).
\label{eq:sndir}
\end{multline}
On March 21 2000 at noon the J2000.0 equatorial coordinate system matches with the geographic one since $\gamma=0^\circ$. For this time Equation~\ref{eq:sndir} is simplified to the same expression used in~\cite{Vedran}. The position of the detector $k$ can be inferred from its latitude ($\phi_k$) and longitude ($\lambda_k$) angles, and the Earth radius (R$_{\rm Earth }$): 
\begin{equation}
    \vec{r}_{k} = R_{\rm Earth} ( \cos\lambda_k \cos\phi_k,\, \sin\lambda_k\cos\phi_k,\,
    \sin\phi_k)
\end{equation}

The centers of the HEALPix pixels~\cite{healpix} with 256 pixels per side (about 0.05\,deg$^2$ per pixel) are used in this work to define the scan grid. The probability that the scanned angles ($\alpha$, $\delta$) coincide with the equatorial coordinates of the CCSN is given by the following $\chi^{2}$ function:
\begin{equation} 
\chi^{2}_{ij}(\alpha,\delta) = 
\left( 
\frac{t_{ij}(\alpha,\delta)-T_{0, ij}^\mathrm{match}}{\delta t_{ij}}
\right)^2 \: ,
\label{eq:chi2tri}
\end{equation}
assuming that there is no systematic shift in the $T_{0, ij}^\mathrm{match}$ determination.
The minimum of the function gives the best estimate for the angles ($\alpha$, $\delta$) of the searched CCSN location in the sky. From Equation~\ref{eq:chi2tri}, one can note that the performance depends on the uncertainty of the measured time delay $\delta t_{ij}$ of each detector pair.

Different detector pairs can be combined into a total $\chi^{2}$ by summing each contribution:
\begin{equation} 
\chi^{2}(\alpha,\delta) = \sum_{i,j}^{i<j} \chi^{2}_{ij}(\alpha,\delta) \: .
\label{eq:chi2tot}
\end{equation}

The $\chi^{2}(\alpha,\delta)$ function is converted into a probability, $p(\alpha,\delta)=p(\chi^2_2 \le \chi^{2}(\alpha,\delta)-\chi^{2}_\mathrm{min})$, which is the cumulative distribution function of the chi-square with two degrees of freedom, $\chi^2_2$, evaluated at the value of chi-square difference between the tested point, $\chi^{2}(\alpha,\delta)$, and the minimum chi-square value, $\chi^{2}_\mathrm{min}$, obtained scanning all possible directions. The 90\% confidence level (C.L.) error region of the source localization is determined as a collection of all points on the sky with a probability $p(\alpha,\delta)<0.9$~\cite{Avni}. 

\subsection{Procedure for the triangulation performance estimation}
\label{s:triangperformance}
The confidence area skymaps can be constructed assigning to each fitted value the true expected delay from a particular CCSN location on the sky and the explosion time, $T_{0,{ij}}^\mathrm{match}=t_{ij}(\alpha_0,\delta_0)$, where $t_{ij}$ is taken from Equation~\ref{eq:tij}. The procedure described in Section~\ref{s:triang} provides the confidence area skymaps which are centred around the true CCSN location with such assumption.

To estimate the size of the confidence area, each fitted value, $T_{0,{ij}}^\mathrm{match}$, is sampled from a normal distribution with the true value, $t_{ij}(\alpha_0,\delta_0)$, as mean and a standard deviation $\delta t_{ij}$. For each simulated set of the delays, $T_{0,{ij}}^\mathrm{match}$, the error box area is calculated at 68 and 90\% C.L. The average values and standard deviations of the error box areas are estimated, repeating the sampling of $T_{0,{ij}}^\mathrm{match}$ sets. The coverage is verified by calculating the fraction of the realisations for which the true CCSN position lies inside the estimated confidence area for each realisation. Additionally, a HEALPix skymap histogram with the density of the fitted source position for all realisations is accumulated and 68 and 90\% C.L. central areas are calculated from it.
\section{Results and performance comparisons}
\label{s:results}
In this section, the estimated resolution of the arrival time delay for different detector pairs is shown. Different combinations of three and four detectors are then used to estimate the resulting CCSN localisation uncertainty area. 

\subsection{Time uncertainty results}
The results of the chi-square method are given for two different light-curve normalisations in Tables~\ref{t:results1} and Table~\ref{t:results2}. For the first normalisation, the true background value is assumed for the baseline subtraction and the scaling is done according the true effective mass. This corresponds to the ideal case in which the detector efficiency and the background estimations are known {\it a priori}. For the second one, the background expectation value is computed as the average rate in a 1\,s off-signal region. The light-curve area in the time window of [--300, 300]\,ms around the light-curve maximum is normalised to one. By comparing Table~\ref{t:results1} and Table~\ref{t:results2} results, the realistic experimental curve normalisation gives similar performances compared to the ideal case. This justifies the proposed window for background estimation and the window used for light-curve normalisation. It is verified that the mean of the $T_0^\mathrm{match}-T_0$ distribution is compatible with zero within the statistical uncertainties. The $\delta t$ obtained from the simulations with a fixed and random delay times are compatible. This confirms that for the assumed model and detector response the chi-square method provides an unbiased estimation of the time delay between the signal arrival at the two detector sites.

\begin{table*}[!h]
\centering
\caption{Uncertainty $\delta t$ in milliseconds obtained with the chi-square method using ideal normalization of the detector neutrino light-curves. The detector pairs are listed in row and column names. The arrow points to the detector name that is used for the light-curve peak definition, which is also shifted during the $\chi^{2}$ scan.}
\label{t:results1}

\resizebox{\linewidth}{!}{
\begin{tabular}{|c|c|c|c|c|}\hline  
&  {\bf KM3NeT-ARCA} & {\bf Super-Kamiokande} & {\bf Hyper-Kamiokande} & {\bf JUNO} \\ \hline
{\bf IceCube}  & $\leftarrow$6.4$\pm$0.2 & $\leftarrow$1.95$\pm$0.04 & $\uparrow$0.53$\pm$0.01 & $\leftarrow$1.95$\pm$0.04 \\ \hline
{\bf KM3NeT/ARCA}  & - & $\uparrow$7.3$\pm$0.2 & $\uparrow$6.5$\pm$0.2 & $\uparrow$7.3$\pm$0.2 \\ \hline
{\bf Super-Kamiokande}  & - & - & - & $\leftarrow$2.73$\pm$0.06 \\ \hline
{\bf Hyper-Kamiokande}  & - & - & - & $\leftarrow$2.02$\pm$0.05 \\ \hline
  \end{tabular}
  }

\end{table*}

\begin{table*}[!h]
\centering
\caption{Uncertainty $\delta t$ in milliseconds obtained with the chi-square method using average background subtraction and unity normalization of the detector neutrino light-curves. The detector pairs are listed in row and column names. The arrows point to the detector name that is used for the light-curve peak definition, which is also shifted during the $\chi^{2}$ scan.}
\label{t:results2}

\resizebox{\linewidth}{!}{
 \begin{tabular}{|c|c|c|c|c|}\hline   
 & {\bf KM3NeT/ARCA} & {\bf Super-Kamiokande} & {\bf Hyper-Kamiokande} & {\bf JUNO} \\ \hline
{\bf IceCube} & $\leftarrow$\,6.65$\pm$0.15 & $\leftarrow$\,1.95$\pm$0.04 & $\uparrow$\,0.55$\pm$0.01 & $\leftarrow$\,1.95$\pm$0.04\\ \hline
{\bf KM3NeT/ARCA} & - & $\uparrow$\,7.4$\pm$0.2  & $\uparrow$\,6.70$\pm$0.15 & $\uparrow$\,7.4$\pm$0.2 \\ \hline
{\bf Super-Kamiokande} & - & - & - & $\leftarrow$\,2.75$\pm$0.06 \\ \hline
{\bf Hyper-Kamiokande} & - & - & - & $\leftarrow$\,1.99$\pm$0.04 \\ \hline
  \end{tabular}
   } 

\end{table*}

\begin{table*}[!h]
\centering
\caption{Uncertainty $\delta t$ in milliseconds obtained with the cross-correlation method using the zero-normalisation. The detector pairs are listed in row and column names. The arrows point to the detector name that is used for the light-curve peak definition.}
\label{t:ccres}

\resizebox{\linewidth}{!}{
 \begin{tabular}{|c|c|c|c|c|}\hline   
 & {\bf KM3NeT/ARCA} & {\bf Super-Kamiokande} & {\bf Hyper-Kamiokande} & {\bf JUNO} \\ \hline
{\bf IceCube} & $\leftarrow$\, 6.2$\pm$0.1 & $\leftarrow$\,2.19$\pm$0.05 & $\uparrow$\,0.64$\pm$0.02 & $\leftarrow$\,2.19$\pm$0.05 \\ \hline
{\bf KM3NeT/ARCA} & - & $\uparrow$\,9.0$\pm$0.2  & $\uparrow$\,6.2$\pm$0.1 & $\uparrow$\,9.0$\pm$0.2 \\ \hline
{\bf Super-Kamiokande} & - & - & - & $\leftarrow$\,5.1$\pm$0.1 \\ \hline
{\bf Hyper-Kamiokande} & - & - & - & $\leftarrow$\,2.59$\pm$0.06 \\ \hline
  \end{tabular}
  }
\end{table*}

For KM3NeT, the effective mass corresponding to the ARCA detector is used for the performance estimation. The maximum signal time delay between the ARCA and ORCA sites, ($\sim$3\,ms), is on the same order of the estimated uncertainties for any of the combinations involving the ARCA detector. This prevents a simple merging of the two KM3NeT light-curves.

The results of the chi-square method shown in Tables~\ref{t:results1} and \ref{t:results2} are computed using the optimized step and bin sizes of 50\,ms. Using 10\,ms for the step and the bin size for ARCA/IceCube and ARCA\-/Hyper-Kamiokande combinations provides slightly better results, reaching the value of 6.20$\pm$0.15 and 6.30$\pm$0.15\,ms, respectively. Adding up to 3~Hz of background rates does not decrease the performance for Super-Kamiok\-ande and JUNO detectors.

The performance of the cross-correlation method using the same search time window as for the chi-square method is shown in Table~\ref{t:ccres}. The optimal effective bin size is found to be 10\,ms for all combinations. The results are compatible with the chi-square method, so cross-correlation represents a viable alternative.

\subsection{Results of the triangulation: localisation skymaps}
Using the uncertainties estimated for each detector combination, the triangulation algorithm is applied to reconstruct the CCSN location on the sky. To estimate the performance of different detector combinations, a CCSN on vernal equinox at noon is assumed for simplicity.  The source direction of the Galactic Centre with equatorial coordinates $(\alpha_0, \delta_0) = (-94.4^{\circ},-28.9^{\circ})$ is chosen as an example.

\begin{figure}[!h]
    \centering
    \includegraphics[width=\fscale\linewidth]{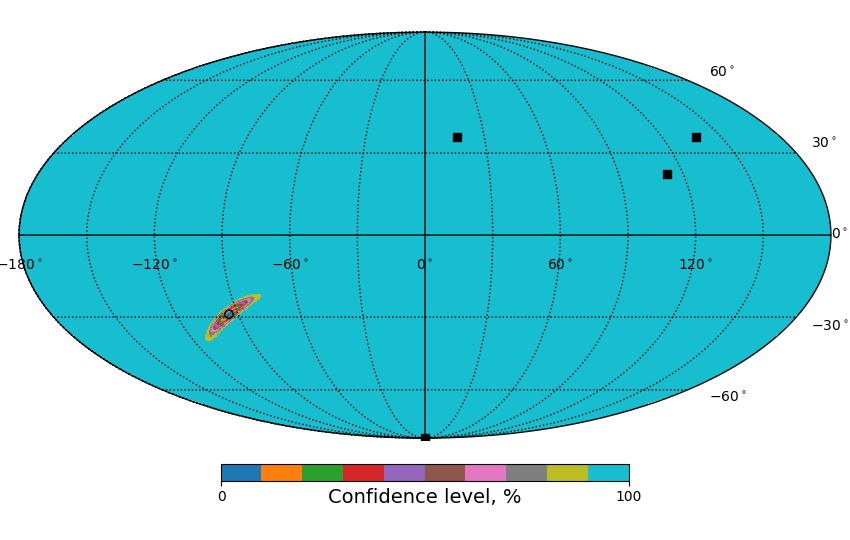}
    \includegraphics[width=\fscale\linewidth]{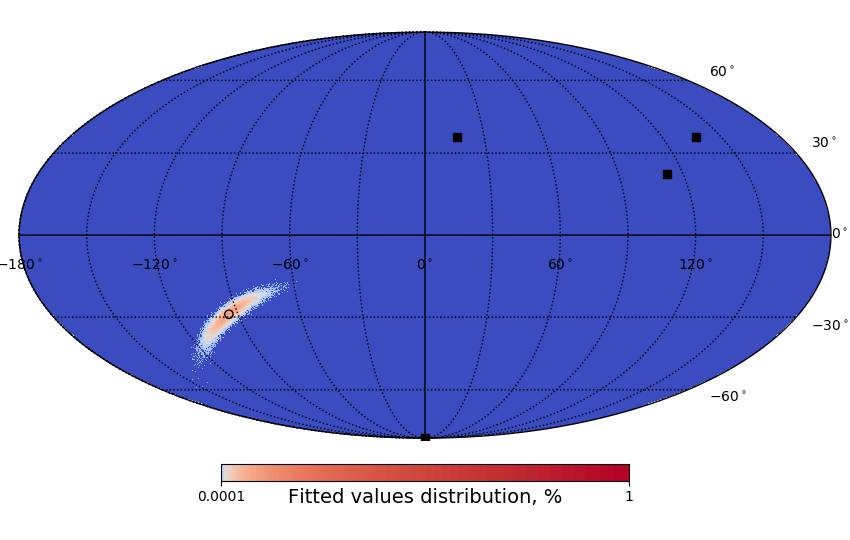}
    
    \caption{Mollweide projection of confidence area in equatorial coordinates for a CCSN at the Galactic Centre (black dot) computed using triangulation between four detectors (black squares): IceCube, KM3NeT/ARCA, Hyper-Kamiokande and JUNO. Top: confidence area assuming true delays, bottom: fitted positions distribution for 100000 realisations of the delay sets.}
    \label{fig:triang4}
\end{figure}

\begin{figure*}[!t]
\includegraphics[width=0.5\linewidth]{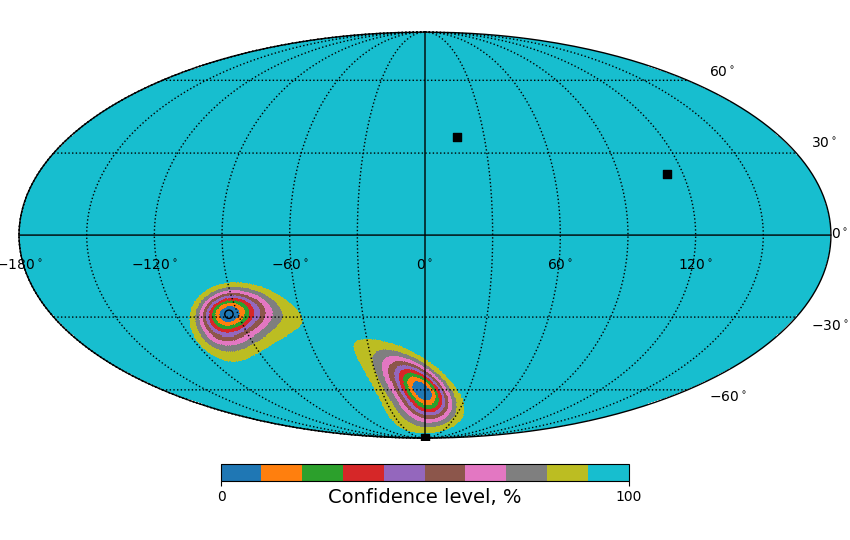}
\includegraphics[width=0.5\linewidth]{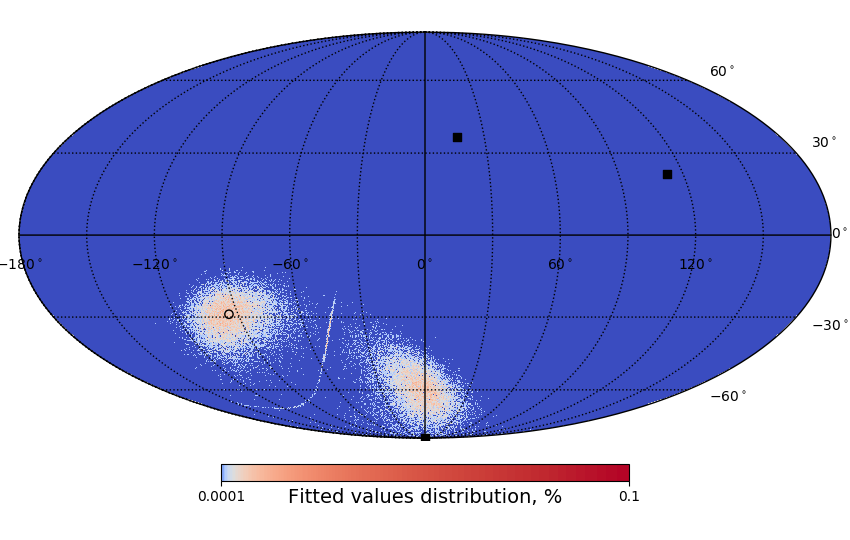}
\includegraphics[width=0.5\linewidth]{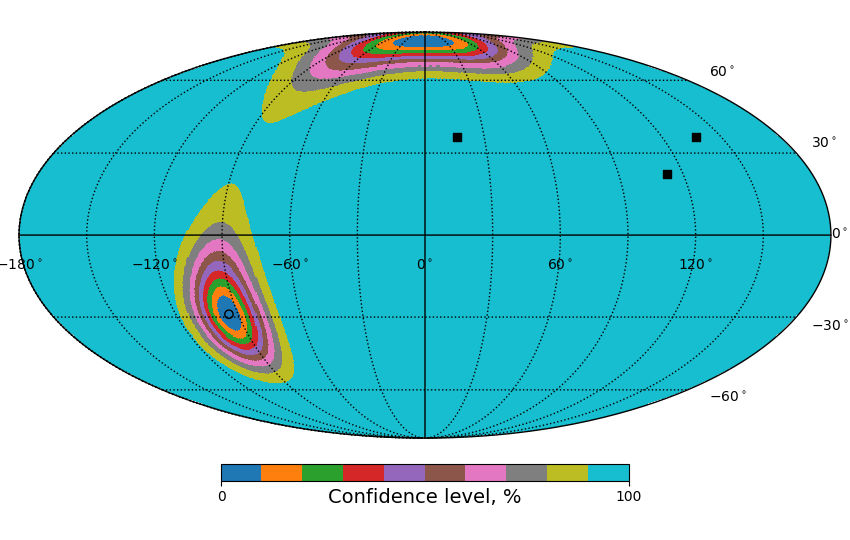}
\includegraphics[width=0.5\linewidth]{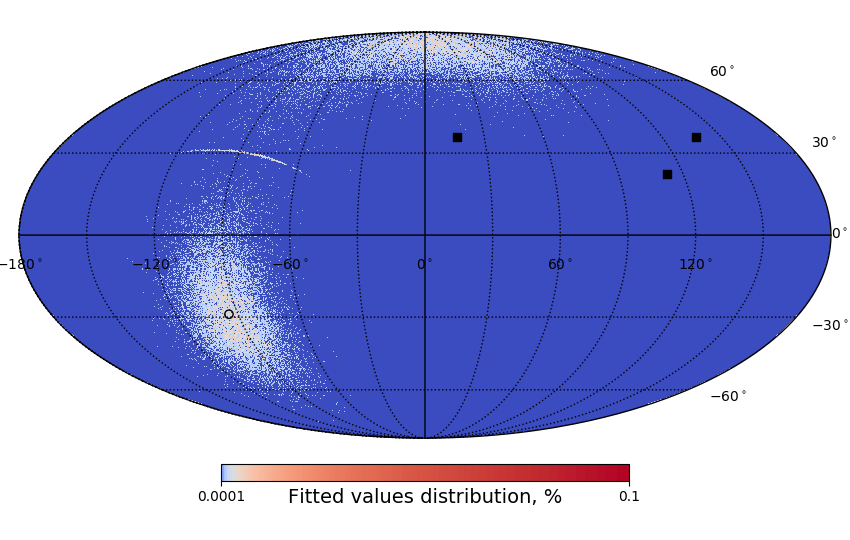}
\includegraphics[width=0.5\linewidth]{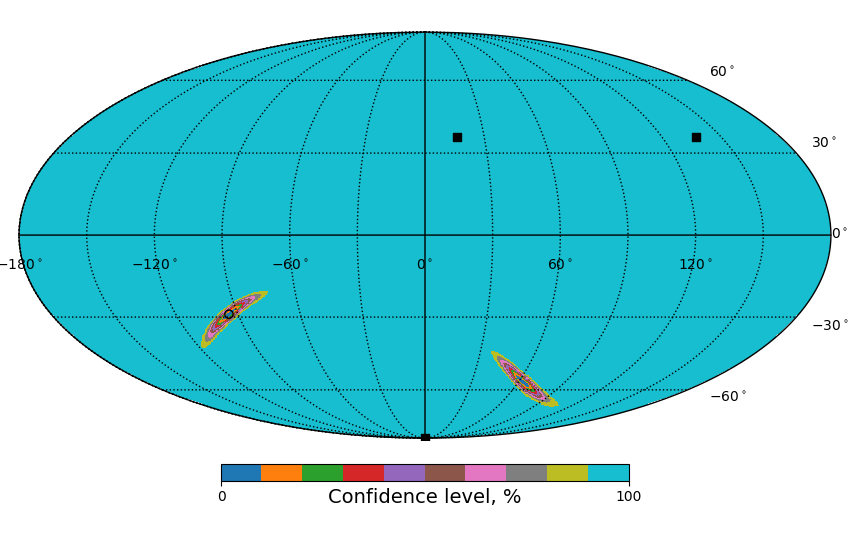}
\includegraphics[width=0.5\linewidth]{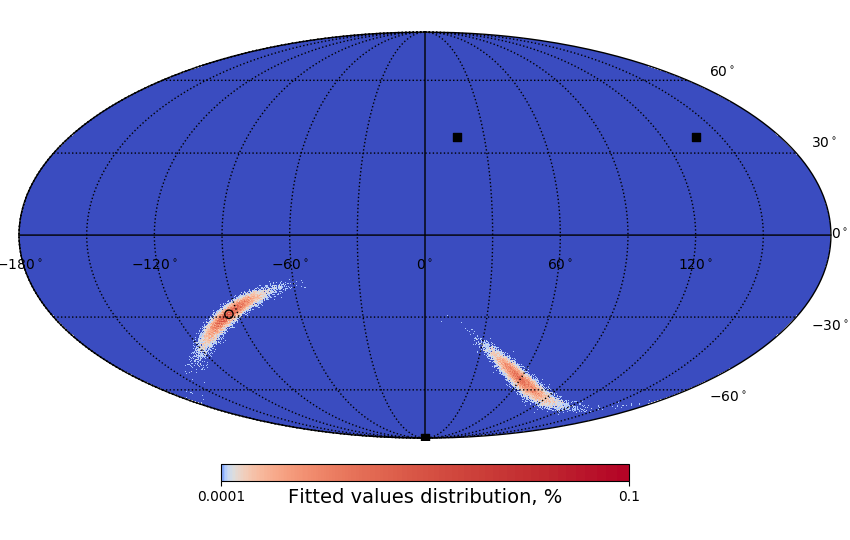}
\includegraphics[width=0.5\linewidth]{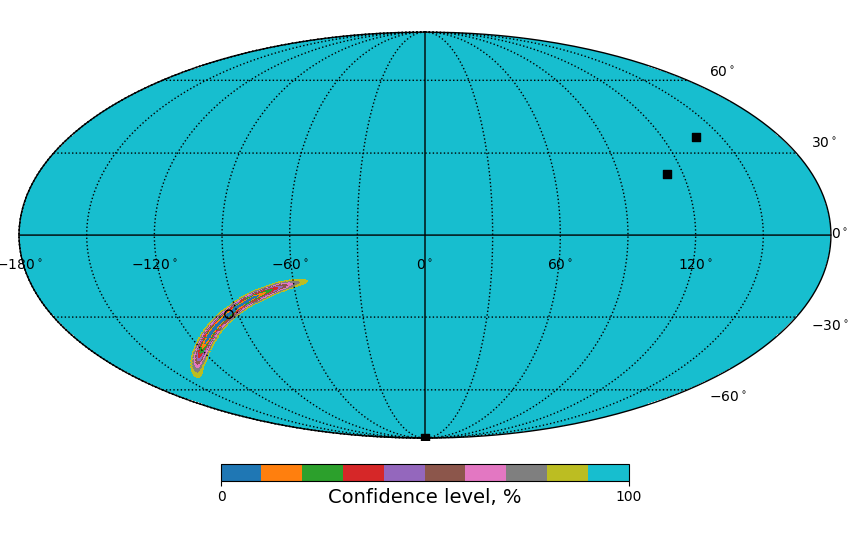}
\includegraphics[width=0.5\linewidth]{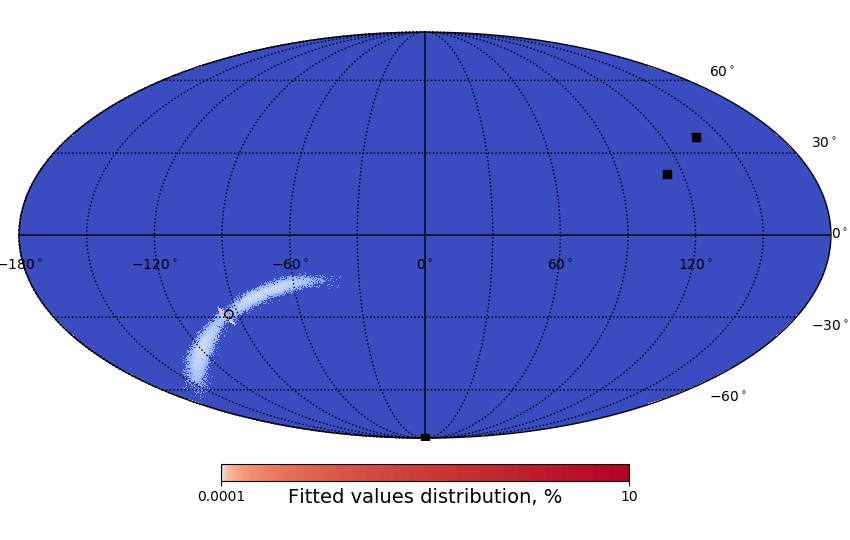}
\caption{Mollweide projection of the confidence areas assuming true delays (left) and fitted position distributions over 100000 realisations (right) in equatorial coordinates for a CCSN at the Galactic Centre (black dot) computed using triangulation between different detector combinations (black squares). From top to down: IceCube, KM3NeT/ARCA and JUNO; Hyper-Kamiokande, KM3NeT/ARCA and JUNO; IceCube, Hyper-Kamiokande and KM3NeT/ARCA; IceCube, Hyper-Kamiokande and JUNO.\\[0.5 cm]}
\label{fig:triang3}
\end{figure*}

The size of each confidence area is obtained with 100000 realisations of $T_{0,{ij}}^\mathrm{match}$ sets. Average values and spreads of the error box sizes at 68 and 90\% C.L. are provided in Table~\ref{t:areas} for different detector combinations in comparison with the areas obtained using the true values for $T_{0,{ij}}^\mathrm{match}$. 
These results are obtained using the uncertainties from the chi-square method given in Table~\ref{t:results2}. The real coverage is verified and the values are provided with statistical uncertainties. The confidence area skymaps shown in Figures~\ref{fig:triang4} and \ref{fig:triang3} are computed assuming the true time delay for each fitted delay. Additionally, the fitted CCSN location distributions for the same realisations are shown in Figures~\ref{fig:triang4}, \ref{fig:triang3} and the sizes of their uncertainty areas are provided in Table~\ref{t:results2}. The real coverage for the error boxes obtained with numerous realisations is higher or at the same level as the set confidence level, so the frequentist coverage is satisfied. Almost in all cases, the obtained average confidence areas are slightly larger compared to the confidence areas obtained assuming the true delays. The uncertainty areas obtained from the fitted position distributions and the ones computed as the average confidence areas are similar when the skymaps have simple elliptical shapes for the confidence areas, but can differ significantly for non elliptical ones and for several disjoint regions.

\begin{table*}[!h]
\centering
\caption{Error box areas of the CCSN localisation in deg$^2$ at 90 and 68\% confidence levels for the considered detector combinations computed in three different ways: a) using the true delays; b) from 100000 randomized time delay realisations; c) as the area covering 90 and 68\% of the fitted position distribution. The real coverage is also provided and it is calculated as a fraction of the realisations (b) for which the true CCSN position lies inside the estimated confidence area for each realisation. The CCSN is assumed to occur at the vernal equinox time in the direction towards the Galactic Centre at a distance of 10 kpc.}
\label{t:areas}
\resizebox{\linewidth}{!}{%
 \begin{tabular}{|c|c|c|c|c|c|c|}\hline
\multicolumn{2}{|c|}{IceCube} & $\checkmark$  &  $\checkmark$ & $\checkmark$ &                         & $\checkmark$ \\
\multicolumn{2}{|c|}{Hyper-Kamiokande} & $\checkmark$  &  $\checkmark$ &                         & $\checkmark$ & $\checkmark$ \\
\multicolumn{2}{|c|}{JUNO}     & $\checkmark$ &                          & $\checkmark$ &  $\checkmark$ & $\checkmark$ \\
\multicolumn{2}{|c|}{KM3NeT/ARCA}    &                         & $\checkmark$  &  $\checkmark$ &  $\checkmark$ & $\checkmark$ \\ \hline 
\multirow{4}{*}{90\% CL}& area with true delays & 350 & 340 & 2060 & 4680 & 140 \\
& average area & 340$\pm$70 & 360$\pm$40 & 2150$\pm$370 & 4680$\pm$660 & 140$\pm$20 \\
& fitted positions area & 230 & 320 & 1440 & 2420 & 130 \\
& real coverage (\%) & 93.3$\pm$0.3 & 90.0$\pm$0.3 & 89.8$\pm$0.3 & 89.9$\pm$0.3 & 90.0$\pm$0.3 \\ \hline

\multirow{4}{*}{68\% CL} & area with true delays  & 200 & 160 & 920 & 2100 & 70 \\
& average area  & 190$\pm$50 & 170$\pm$20 & 1050$\pm$230 & 2300$\pm$460 & 70$\pm$10 \\
& fitted positions area &  70 & 160 & 720 & 1270 & 70 \\
& real coverage (\%) & 77.3$\pm$0.3 & 67.8$\pm$0.3 & 68.0$\pm$0.3 & 68.2$\pm$0.3 & 68.2$\pm$0.3 \\ \hline
  \end{tabular}
 }
\end{table*}

The results in Table~\ref{t:areas} indicate that a favourable position of a detector with respect to the source location and other detectors may compensate for worse time resolution, for example, when JUNO is replaced with KM3NeT/ARCA in the combination of three detectors. Note, however, that for the considered CCSN location and time, the three detector combinations involving KM3NeT/ARCA provide confidence areas with two disjoint regions. The confidence areas may be further reduced considering their intersection with the Galactic Plane.

To verify if the triangulation precision depends on the position of the CCSN on the sky, two more directions were tested: the one from the Betelgeuse star $(88.8^{\circ},7.4^{\circ})$ as the currently most promising progenitor and another, compatible with the Cygnus constellation $(-45.0^{\circ},40.0^{\circ})$, since it is located on the opposite hemisphere respect to the Galactic Centre. For these two additional directions the areas of the error region at 1~$\sigma$ C.L. are 53$\pm$4\,deg$^{2}$ and 50$\pm$4\,deg$^{2}$, respectively, which is on same order of magnitude as for the direction towards the Galactic Centre (70$\pm$10\,deg$^2$). The expected time delay for each detector pair is given in Table.~\ref{t:offsets} for the three considered sources, together with the uncertainty on its estimation with the chi-square method.

\begin{table*}[!h]

\centering
\caption{Time delay for each detector pair, assuming the signal emission coming from the three different sources. The time delay uncertainty for each detector pair from Table~\ref{t:results2} is given in the last column.}
\label{t:offsets}
\begin{tabular}{|c|c|c|c|c|}\hline  
& {\bf Galactic Centre} & {\bf Betelgeuse} & {\bf Cygnus} & { $\boldsymbol{\delta t}$} \\ \hline
{\bf IceCube -- Hyper-Kamiokande}     & -25.8\,ms & 15.7\,ms  & 8.7\,ms  & 0.55$\pm$0.01\,ms \\ \hline
{\bf IceCube -- KM3NeT/ARCA}          & -21.7\,ms & 9.4\,ms   & 28.2\,ms & 6.65$\pm$0.15\,ms \\ \hline
{\bf IceCube -- JUNO}                 & -29.6\,ms & 21.7\,ms  & 4.9\,ms  & 1.95$\pm$0.04\,ms \\ \hline
{\bf Hyper-Kamiokande -- KM3NeT/ARCA} & 4.1\,ms   & -6.2\,ms  & 19.6\,ms & 6.70$\pm$0.15\,ms \\ \hline 
{\bf Hyper-Kamiokande -- JUNO}        & -3.9\,ms  & 6.1\,ms   & -3.8\,ms & 1.99$\pm$0.04\,ms \\ \hline
{\bf  KM3NeT/ARCA -- JUNO}             & -8.0\,ms   & 12.3\,ms & -23.3\,ms & 7.4$\pm$0.2\,ms\\ \hline
\end{tabular}

\end{table*}
The results of our work can be compared to the latest triangulation studies. In~\cite{Vedran}, the estimate of the 68\% C.L. area is $\sim$66~deg$^{2}$. This result is similar compared to the one obtained in our work (70$\pm$10\,deg$^{2}$), although more than four detectors were combined using IBD and ES channels and the uncertainty estimate relies on the matching with a light-curve template known {\it a priori}. The result of our work represents a model independent data analysis proposal. In~\cite{Hansen} the timing with the first IBD events has the best performance comparing to the exponential rise fit. In the latest results with the first events observation method~\cite{Kate}, the time delay uncertainty for Super-Kamiokande and JUNO combination is 5.7\,ms. This is larger than 2.8\,ms estimated in our work, however, a more pessimistic model was used in the former work. For a four times lower flux in a simplified model used here this uncertainty degrades to $6.3\pm0.2$\,ms. The method in~\cite{Kate} requires an evaluation and correction for several biases due to background rates and the steepness of the luminosity curve rise. Biases and the performance degradation in the method proposed here may appear due to the detector efficiency varying with the neutrino energy and in time, or with respect to the event rates. A proper estimation of such effects is possible considering simulations with a detailed emission model and precise detector parameters. This new method relies on the agreement among the different collaborations for making their full light-curve available to SNEWS while the method in~\cite{Kate} only requires sending the time information of the first events. 

The results of this work can also be compared with the expected performance of Super-Kamiokande using the directionality information from the elastic scattering channel. The 68\% C.L. area for the combination of four detectors in this work is similar to the area expected with the actual Super-Kamiokande configuration ($\sim$69~deg$^{2}$~\cite{SKpointing}). With gadolinium doping, this area might be reduced down to $\sim$13~deg$^{2}$~\cite{SKpointing}. The expected CCSN 68\% C.L. area for the JUNO IBD events reconstruction analysis will be better than 254\,deg$^2$~\cite{Juno}. The triangulation method can be proposed as a low latency analysis. The confidence areas from Super-Kamiokande, JUNO and this triangulation analysis are independent and can be combined in order to obtain a joint refined measurement.
\section{Conclusion}
\label{s:concl}
Detectors in current and future operation will be capable of collecting a significant number of neutrinos from the next galactic CCSN explosion. This will allow for detailed studies of the time profile. The determination of the neutrino arrival time is crucial for the source localisation that may help for a potential multi-messenger follow-up.

The detected neutrino light-curves can be used to deduce the delay in the arrival time of the supernova emission at different sites on Earth. This paper describes the chi-square and cross-correlation methods used to estimate such delays and their uncertainties. Since a direct comparison of the experimental curves is used, these methods do not rely on any model for the alert algorithm implementation.

Supernova emission consists of a neutrino flux with a different light-curve for each flavour. Therefore, the compared experimental light-curves should consist of events detected through the same neutrino interaction process. This is the case for most detectors sensitive to the electron anti-neutrino interactions via inverse beta decay in water. The method can be extended to detectors sensitive to different channels accounting for systematic biases, which can be estimated on a model dependent basis.

The methods are tested using a time dependent parametrisation of the neutrino luminosity. The detectors are described with two parameters, the effective mass and the background rate. The signal and background are sampled assuming Poisson distributions. These simulations may be further improved by the respective collaborations using accurate detector simulations. Detailed CCSN fluxes can be used in the simulation to estimate the performance of the methods for any given model.

Merging the time delay information between several detector sites allows to infer the supernova localisation. The 90\% confidence area is 140$\pm$20\,deg$^2$ when combining Hyper-Kamiokande, IceCube, JUNO and KM3NeT/ARCA detectors. Such analysis can be performed in real-time within the framework of the SNEWS alert system. This location uncertainty can be reduced further by intersecting this area with the CCSN progenitors distribution in the Milky Way and combining with the confidence areas from the  Super-Kamiokande and JUNO detectors performing stand-alone source localisation.

Using a bootstrapping strategy, the algorithm parameters can be optimised directly on data to improve the time resolution and to localise the source with better accuracy.
\section{Supplementary materials}
The methods described in this work are implemented and stored in a public repository~\cite{codes}. It consists of the following packages:
\begin{itemize}
\item simulation: detected neutrino light curve simulation codes (C++),
\item matching: chi-square and cross-correlation matching codes (C++),
\item skymap: HEALPix \cite{healpix} equatorial skymap creation code and error box area calculation (Python).
\end{itemize}

\section{Acknowledgements}
The authors would like to thank Alec Habig and Vedran Brdar for the fruitful discussion at the Orsay workshop on the supernova detection and the further support, Kate Scholberg from the SNEWS project, Lutz K\"{o}pke, Erin O'Sullivan and Segev BenZvi from the IceCube Collaboration and the KM3NeT Collaboration for the interest in this activity. This project is  financially supported by the LabEx UnivEarthS (ANR-10-LABX-0023 and ANR-18-IDEX-0001). This project has received funding from the European Union’s Horizon 2020 research and innovation programme under grant agreement No 739560.


\end{document}